\documentstyle[12pt]{article}

\def\eq#1{Eq.~(\ref{#1})}
\def\eqs#1#2{Eqs.~(\ref{#1}) and (\ref{#2})}

\def\rfrac#1#2{\left(\frac{#1}{#2}\right)}

\def\text#1{\hbox{\rm#1}}

\def\pr{^\prime}

\def\ee{\end{equation}}
\def\be{\begin{equation}}
\def\eea{\end{eqnarray}}
\def\bea{\begin{eqnarray}}
\def\eqa{\!\!\!&=&\!\!\!}
\def\equiva{\!\!\!&\equiv &\!\!\!}

\def\simeqa{\!\!\!&\simeq &\!\!\!}

\def\lla{\!\!\!&\ll &\!\!\!}

\def\GeV{\,\text{GeV}}

\def\km{\,\text{km}}

\def\Mpc{\,\text{Mpc}}

\def\mone{^{-1}}

\def\mhalf{^{-1/2}}

\def\half{^{1/2}}

\def\twothird{^{2/3}}

\def\quarter{^{1/4}}

\def\pr{^\prime}

\def\s{\text s}

\def\E{\text E}

\def\bfa#1{{\bf #1}}

\def\lsim{\mathrel{\rlap{\lower4pt\hbox{\hskip1pt$\sim$}}
    \raise1pt\hbox{$<$}}}
\def\gsim{\mathrel{\rlap{\lower4pt\hbox{\hskip1pt$\sim$}}
    \raise1pt\hbox{$>$}}}

\def\calp{{\cal P}}

\begin{document}
  \begin{titlepage}
  \begin{flushright}
SUSSEX-AST 92/6-1; LANC-TH 5-92\\
(June 1992)\\
  \end{flushright}
  \begin{center}
{\Large
{\bf COBE, Gravitational Waves, Inflation and Extended Inflation\\}}
\vspace{.3in}
{\large Andrew R.~Liddle$^{\dagger}$ and David H.~Lyth$^*$\\}
\vspace{.4 cm}
{\em  $^{\dagger}$Astronomy Centre, \\ Division of Physics and Astronomy, \\
University of Sussex, \\ Brighton BN1 9QH.~~~U.~K.}\\
\vspace{.4 cm}
{\em  $^*$School of Physics and Materials,\\ Lancaster University,\\
Lancaster LA1 4YB.~~~U.~K.}\\
\end{center}
\vspace{.2 cm}
\baselineskip=24pt
\begin{abstract}
\vspace*{1em}\noindent
We analyse the implications for inflationary models of the cosmic microwave
background (cmb) anisotropy measured by COBE. Vacuum fluctuations during
inflation generate an adiabatic density perturbation, and also gravitational
waves. The ratio of these two contributions to the cmb anisotropy is given for
an arbitrary slow-roll inflaton potential. Results from the IRAS/QDOT and
POTENT galaxy surveys are used to normalise the spectrum of the density
perturbation on the scale $20h^{-1}\Mpc$, so that the COBE measurement on the
scale $10^3h^{-1}\Mpc$ provides a lower bound on the spectral index $n$. For
`power law' and `extended' inflation, gravitational waves are significant and
the bound is $n>0.84$ at the $2$-sigma level. For `natural' inflation,
gravitational waves are negligible and the constraint is weakened to $n>0.70$,
at best marginally consistent with a recent proposal for explaining the excess
clustering observed in the APM galaxy survey. Many versions of extended
inflation, including those based on the Brans--Dicke theory, are ruled out,
because they require $n\lsim 0.75$ in order that bubbles formed at the end of
inflation should not be observed now in the cmb.
\end{abstract}

E-mail addresses: arl @ uk.ac.sussex.starlink~~~;~~~d.lyth @ uk.ac.lancaster

\vspace{0.2cm}
\begin{center}
{\em Final version, to appear, Physics Letters B}
\end{center}
\end{titlepage}

In this note we discuss the implications of the COBE measurement of the
aniso\-tropy of the cosmic microwave background \cite{S92}. First we consider
only the contribution of the density perturbation (Sachs--Wolfe effect). The
spectrum of the density perturbation is normalised on the scale $20h\mone\Mpc$
using results from the IRAS/QDOT and POTENT galaxy surveys. The COBE
measurements constrain the spectrum on the scale $10^3h^{-1}\Mpc$, and
therefore constrain the index $1-n$ of its assumed power-law scale dependence.
We then include the effect of gravitational waves, calculated along with the
Sachs-Wolfe effect in various inflationary models. Our most dramatic result is
that many versions of extended inflation seem to be ruled out.

We define the spectrum of the matter density contrast $\delta =
\delta\rho/\rho$, evaluated at the present epoch, as $\calp_\rho(k) =
V(k^3/2\pi^2) \langle|\delta_\bfa k|^2\rangle$ where $\delta_{\bf k} =
V\mone\int e^{-ik.x} \delta(x) d^3x$ is the Fourier coefficient in a box of
volume $V$ and the brackets denote the average over a small region of
$k$-space. The density contrast, averaged over a  randomly placed sphere of
radius $R$, has a mean square given by
\be \sigma^2(R)=\int^\infty_0 W^2(kR) \calp_\rho(k)
\frac{dk}{k} \ee
where the window function $W$ is given in \cite{PEEBLES,KOT90}. The scale
$k\mone\sim R$ dominates the integral.

On scales $k\mone\gsim 10h\mone\Mpc$ the density contrast is still evolved
linearly. (As usual $h$ is the Hubble parameter $H_0$ in units of
$100\km\,\s\mone\Mpc\mone$, and we take  $h=.5$.) Its spectrum is of the form
\cite{PEEBLES,KOT90},
\be \calp_\rho= \delta^2_H(k) T^2(k) \rfrac{k}{H_0}^4 \label{two} \ee
Up to a gauge dependent numerical factor the primeval spectrum is of the same
form with $T=1$, so it is specified by $\delta_H^2$. The transfer function
$T(k)$ is practically equal to 1 on scales $k\mone\gg 100\Mpc$. On smaller
scales it depends on the nature of the dark matter, but for the scales
$k\mone>20h\mone\Mpc$ that we are considering the dependence is weak. We use
the cold dark matter transfer function parametrised in \cite{E90}.

We assume that on scales of interest
there is a power law spectrum,
$\delta_H^2\propto k^{n-1}$
for some spectral index $n$,
and ask what departures there may be from the scale-invariant
value $n=1$ advocated by Harrison and Zel'dovich.

\vspace*{1em}\noindent\it
1. Normalisation at $20 h\mone \Mpc$. \\
\rm
Following Efstathiou, Bond and White \cite{EBW} and Taylor and Rowan-Robinson
\cite{TR} in their treatment of the COBE data, we normalise the density
contrast $\delta N/N$ on a scale of roughly $20 h\mone\Mpc$ using the
QDOT/IRAS galaxy map \cite{IRAS} and the POTENT peculiar velocity map
\cite{BD}. The average number density contrast of IRAS galaxies in a randomly
placed cube with sides of $30h^{-1}\Mpc$ has rms \cite{SRL} $0.46 \pm 0.07$.
We have doubled the quoted uncertainty, corresponding presumably to the
$2$-sigma level which we adopt throughout. To simplify the subsequent
analysis, we pretend that the result applies to a sphere of the same volume as
the cube, ie., to a sphere of radius $19h\mone\Mpc$.

This result is converted into a result for the mass density contrast $\sigma$,
on the assumption that $\delta N/N = b_I\delta\rho/\rho$, where the IRAS bias
factor $b_I$ is assumed to be position independent. According to \cite{TR},
three independent dynamical estimates of $b_I$ give respectively $1.23 \pm
0.23$, $1.16 \pm 0.23$ and $1.2 \pm 0.1$. As they use the same data the errors
on these numbers cannot be combined statistically, but it seems reasonable to
take as something like a $2$-sigma result $b_I= 1.2 \pm 0.3$. This gives
\be \sigma(19h\mone \Mpc)=0.38 \pm 0.11 \label{five}\ee

For ease of comparison with other work, one specifies the normalisation by
giving the bias factor $b_8\equiv\sigma\mone(8h\mone\Mpc)$, where $\sigma$ is
the linearly evolved quantity calculated with the cold dark matter transfer
function. We emphasise that this is just a way of giving the result, the
present paper assuming nothing about what goes on below the $20h\mone\Mpc$
scale. The result corresponding to \eq{five} can be accurately parametrised as
\be b_8 = (1.1\pm .3) \rfrac{2.9-n}{1.9}\twothird \label{six} \ee
(cf a similar result in \cite{EBW}).

\vspace*{1em}\noindent\it
2. The effect of the density perturbation on COBE. \\
\rm
The cosmic microwave background anisotropy observed by COBE probes scales
$\gsim 10^3\Mpc$, corresponding to the angular resolution $\sim 10^0$. Because
of the `cosmic variance' problem mentioned later, we focus on the lower end of
this range. Specifically, we use the result that the anisotropy $\delta T/T$
(excluding the dipole) smeared by a $10^0$ full-width at half-maximum Gaussian
window has the rms value \cite{S92} $\sigma_T(10^0) = (1.1 \pm 0.4) \times
10^{-5} $. We have doubled the quoted uncertainty to obtain a $2$-sigma
quantity. In terms of the multipoles $a_{\ell m}$ defined by $\delta T/T=\sum
a_{\ell m} Y_{\ell m}(\theta,\phi)$, one has $\sigma_T^2(10^0) =
2\sum_{\ell=2}^\infty A_\ell^2 F_\ell$, where $A_\ell^2=\sum_m |a_{\ell
m}|^2/4\pi$ and the filter function is \cite{E90,BE}$F_\ell=.5\exp[
(-4.25\pi\ell/180)^2 ]$.
Cosmological perturbation theory gives
the expectation value of $A_\ell^2$ for a randomly placed observer in
the form
\be
\langle A_\ell^2 \rangle =\langle A_\ell^2|_\rho \rangle
+\langle A_\ell^2|_g\rangle
\ee
where the first term is the contribution of the density contrast (Sachs--Wolfe
effect) and the second term is the contribution of gravitational waves.%
\footnote
{In both this expression and in \eq{two} for $\calp_\rho$ we are neglecting
any contributions from a primeval isocurvature density perturbation, strings,
textures  and so on. Any such contribution will strengthen our conclusion if
it contributes more to the cmb anisotropy than it does to the density
perturbation on the scale $20h\mone\Mpc$, and weaken it if the converse is
true.}
The first term is \cite{PEEB82}
\be
\langle A_\ell^2|_\rho \rangle
=\frac{2\ell+1}{4}\int^\infty_0 j_\ell^2(2k/H_0)\delta_H^2(k) \frac
{dk}{k} \ee
For a power-law spectrum it becomes \cite{SV,SAL}
\be <A_\ell^2|_\rho>=\frac18 \left[\frac{\sqrt\pi}{2}\ell(\ell+1)
\frac{\Gamma((3-n)/2)}{\Gamma((4-n)/2)}
\frac{\Gamma(\ell+(n-1)/2)}{\Gamma(\ell+(5-n)/2)}
\right]
\frac{2\ell+1}{\ell(\ell+1)}\delta^2_H(H_0/2)
\label{first}
\ee
For $n=1$ the square bracket is equal to 1. For $\ell\gg1$ and $\ell\gg |n|$
it can be replaced by 1 if $\delta_H$ is evaluated on the scale $k\simeq \ell
H_0/2$ which dominates the integral.

The variance (mean square deviation) from this expectation value, giving the
uncertainty on the predicted result for our particular position, is called the
cosmic variance. It can be calculated on the assumption that the probability
distribution of each multipole is Gaussian (which is the case for the
inflationary models considered here), and it is large for low multipoles,
falling rapidly as $\ell$ increases. One can readily calculate, assuming the
probabilities are independent, that the cosmic variance for fluctuations on
$10^0$ is only 10\% \cite{S92,SV}, which when combined in quadrature with the
observational rms uncertainty increases it from $17\%$ to $20\%$. We ignore
the cosmic variance here, but note that it will dominate the observational
uncertainty when the latter is reduced by more than a factor of 2.

Using \eq{first} we have calculated the density contribution to the $10^0$
anisotropy, and deduced the normalisation of its spectrum on the assumption
that other contributions are negligible. Expressed in terms of $b_8$ the
result is
\be b_8=(.95^{+.55}_{-.25}) \exp[2.62(1-n)] \label{seve} \ee
where as always the uncertainties are $2$-sigma. Combining this result with
the IRAS normalisation \eq{six} determines $n$. The central value is close to
one, and the range allowed by stretching each normalisation to its $2$-sigma
limit is $0.70<n<1.27$. This result may be compared with the one derived by
the COBE group from their data alone, which at the $1$-sigma level is
$n=1.1\pm.5$. The weakness of the latter result is due partly to the fact that
only one decade of scales ($10^3\Mpc$ to the horizon) is used instead of the
two decades used to obtain the former result, and partly to the large cosmic
variance in the upper part of this decade.

In an earlier publication \cite{LLS} we advocated a value $n \simeq 0.6$ to
fit the APM galaxy correlation function. The bound just derived is at best
marginally compatible with such a proposal, and we have yet to consider the
effect of gravitational waves.

\vspace*{1em}\noindent \it 3. Inflation and gravitational waves. \\ \rm
Given an inflaton potential which leads to slow-roll inflation, one can
calculate the spectrum $\calp_g$ of the gravitational waves, as well as the
spectrum $\calp_\rho$ of the density perturbation, and hence determine the
ratio
\be R_\ell\equiv \langle A_\ell|_g^2\rangle/
\langle A_\ell|_\rho^2\rangle \ee
This is to be evaluated for the relevant scale $k\simeq \ell H_0 /2$. The
ratio depends only weakly on $\ell$ in the range $\ell\lsim 10$ of interest
here, for the inflationary models which we shall discuss. We shall therefore
drop the subscript $\ell$.

The effect of $R$ is to multiply the upper bound on the density perturbation
implied by a given cmb anisotropy by a factor $(1+ R)\mhalf$, which multiplies
\eq{seve} by a factor $(1+R)\half$ and tightens the bound on the spectral
index $n$. This has recently been emphasised in connection with the COBE data
by Salopek \cite{SAL}, who calculated the effect of gravitational waves
numerically. Here we give a simple analytic formula for $R$, and also the
spectral index $n$, in terms of the inflaton potential and its first two
derivatives. We go on to apply them to models of inflation, including extended
inflation.

One generally assumes that the inflaton field $\phi$ and its potential $V$
satisfy the following `slow-roll' conditions%
\footnote{For power-law inflation exact results will be mentioned later.},
at the epoch when relevant scales leave the horizon during inflation
\cite{LINDE,KOT90}
\bea \dot\phi\eqa-\frac{V\pr}{3H}
\label{fotea} \\
\epsilon \lla 1
\label{foteb} \\
|\eta|\lla 1
\label{fotec}
\eea
where
\bea
\epsilon\equiva \frac{m_P^2}{16\pi}\rfrac{V\pr}{V}^2\simeq
-\frac{\dot H}{H^2}\\
\eta\equiva \frac{m_P^2}{8\pi}\frac{V^{\prime\prime}}{V}
\eea

Here $H$ is the Hubble parameter during inflation, equal to $\dot a/a$ where
$a$ is the scale factor of the universe with present value $a=1$. We will
denote the epoch when inflation ends by a subscript 2, and the epoch $k=aH$
when the scale $k\mone$ leaves the horizon by a subscript $k$. They may be
related by equating two expressions for the number $N(t_k)$ of Hubble times
before the end of inflation (eg.~\cite{LYT90})
\be N(t_k)=\frac{8\pi}{m_P^2} \int^{\phi_2}_{\phi_k}
\frac{V}{V\pr} d\phi
=60+\ln\frac{V_k\quarter}{10^{16}\GeV}
+2\ln\frac{V_k\quarter}{V_2\quarter}-\ln(k/H_0) \label{site} \ee
(The second expression assumes that reheating is prompt, otherwise there is a
small correction.) Except in the case of power-law inflation, the end of
inflation is supposed to be signalled by the failure of \eq{foteb} or
\eq{fotec},
\be \max\{\epsilon_2,|\eta_2|\}=1 \label{sete} \ee

The spectrum of the adiabatic density perturbation is given by
\cite{DHL85}
\be \delta_H^2(k) =\frac{32}{75}
\frac{V_k}{m_P^4} \epsilon_k\mone \label{eigh} \ee
Using \eq{fotea}, the effective spectral index on a given scale is therefore
given by
\be 1-n=-\frac{d\ln [\delta_H^2(k)]}{d\ln k}
	= \left(6\epsilon_k-2\eta_k\right)
\label{fite} \ee
The spectrum of the
canonically normalised gravitational wave amplitude is the same as that of any
other massless scalar field
\cite{VF}
\be \calp_g(k)\simeq\rfrac{H_k}{2\pi}^2\simeq\frac2{3\pi}\frac{V_k}{m_P^2}
\label{nine} \ee
Its effective spectral index is given by
\be n_g=-\frac{d\ln \calp_g}{d\ln k}
= 2\epsilon_k \label{grin} \ee

When it is small, the correction to \eqs{eigh}{nine} is of order
$\max\{\epsilon_k,|\eta_k|\}$ \cite{LS2}, which is usually of order $|1-n|$.

Using \eq{first} on the appropriate scale, the density contribution to the cmb
anisotropy is
\be \frac{\ell(\ell+1)}
{2\ell+1} \langle A_\ell^2|_\rho \rangle \simeq
\frac{4}{75} \frac{V_k}{m_P^4}\epsilon_k\mone
\label{ten} \ee
The gravitational contribution is given by
\cite{STAR}
\be \frac{\ell(\ell+1)}
{2\ell+1} \langle A_\ell^2|_g \rangle \simeq
\frac{8}{27} \left( 1+\frac{48\pi^2}{385} \right) C_\ell
\frac{V_k}{m_P^4} \label{elev} \ee
In these expressions $k\simeq \ell H_0/2$, the scale dominating the $\ell$th
multipole, and the correction factor $C_\ell$ in the second expression is
equal to 1 for $\ell\gg 1$, with $C_2=1.1$. They lead to the ratio
\be R=12.4C_\ell \epsilon_k\simeq 12.4 \epsilon_k
\label{thte} \ee
Since $\epsilon=-\dot H/H^2$, we see that the gravitational contribution is
negligible if and only if inflation is almost exponential.

\vspace*{1em}\noindent \it 4. Inflationary models. \\
\rm
Let us use these expressions to calculate $R$, $n_g$ and $n$, for the usually
considered types of inflaton potential. We need them only for the few Hubble
times following the epoch when the scale $k=H_0$ leaves the horizon, which we
will denote by a subscript 1, which for the potentials considered here means
that they are practically scale independent.

`Chaotic' inflation \cite{LINDE} employs a potential $V\propto\phi^{\alpha}$.
{}From \eqs{site}{sete} one learns that $\phi_1^2\simeq120\alpha m_P^2/8\pi$.
This leads to $1-n=(2+\alpha)/120$  and $R=.05\alpha$, in excellent agreement
with Salopek's numerical results \cite{SAL}. With the usual choices $\phi^2$
and $\phi^4$, $n$ is very close to 1. The ratio $R$ is respectively $.1$ and
$.2$, which is similar to the cosmic variance with a $10^0$ resolution. One
therefore needs a better resolution to detect the gravitational waves in this
model.

Power-law inflation \cite{AW,LM} employs a potential
\be V\propto \exp\left(\sqrt\frac{16\pi}{p}\frac{\phi}{m_P}
\right)\label{thtea} \ee
corresponding to  a scale factor $a\propto t^p$ where $t$ is the time during
inflation. From \eqs{fite}{grin}, the spectral indices
are given by $n_g=1-n=2/p$; they are independent of scale so that
the spectra have a genuine power law dependence.
The ratio $R$
is equal to $\simeq12.4/p$.
The value of $Rp/12.5$ for $\ell=2$ can be read off Figure~2 of
\cite{FLM}. It is within $10\%$ or so of unity for $.6<n<1$, so in this range
we feel confident in setting $R=12/p$ for all $\ell$. As shown in Figure~1,
this leads to the bound $n>0.84$, at the $2$-sigma level separately on each
piece of data. (In calculating these curves we have used the
exact expressions $n_g=1-n=2/(p-1)$ \cite{AW,LM,LS}.)

The inclusion of gravitational waves has tightened the bound on $1-n$ by a
factor of 2, and made it impossible to fit the APM data! A similar conclusion
was reached recently in a more qualitative fashion by Salopek \cite{SAL}. Note
also how rapidly this bound improves if the true result is not at the top of
the COBE $2$-sigma range. For instance, were the COBE result to prove exactly
right then the bound would be around $n > 0.9$.

It is, however, possible to generate a power-law spectrum with negligible
gravitational radiation; this is achieved by the `natural' inflation
\cite{FFO} potential
\be V\propto[1+\cos(\phi/f)] \ee
\eqs{site}{sete} determine $\phi_1/f$
We assume that $f<m_P$, and in this regime $\phi_1/f\ll \pi$ so that
\bea 1-n\simeqa \frac{m_P^2}{8\pi}\frac1{f^2} \\
R\simeqa \frac3{16\pi} \frac{m_P^2 \phi^2}{f^4} \eea
With $f=m_P$ one has $n=.96$ and $R=.02$. As $f$ decreases, $R$ decreases very
rapidly, for instance $R\sim 10^{-5}$ for $f=.5 m_P$. Thus $R$ is indeed
negligible. At the minimum allowed value $f=.3 m_P$, $n=.5$ so it seems
possible to have $n\simeq.6$ in this model. Note however that the corrections
to \eqs{eigh}{ten} have yet to be calculated, in contrast with the power-law
case.

The normalisation of a given inflationary potential is determined by the
normalisation of $\calp_\rho$, for instance \eq{six}. In all cases one has
$V_2\quarter<V_1\quarter\lsim 10^{16}\GeV$, where $k\sim H_0$ is the largest
relevant scale \cite{LYTH84,LYT90,LS,LLS}. For the following discussion of
extended inflation we will need an upper limit on $V_2\quarter$ in the case of
power-law inflation. We have calculated it using the exact results of
\cite{LS} with the $2$-sigma limit of \eq{six}, and the result is plotted in
Figure~2.

\vspace*{1em}\noindent\it 5. Extended inflation. \\
\rm
Extended inflation \cite{LS89,K91} reintroduces the original concept
\cite{GUTH} of inflation driven by a first-order phase transition, brought to
an end by bubble nucleation. The Lagrangian to which it corresponds can be
written either in the `Jordan frame' where the matter terms are canonical or
the `Einstein frame' where gravity is canonical, the two `frames' being
related by a conformal transformation of the metric. The Jordan frame is the
most convenient after inflation, and also during inflation when one is
considering an ordinary scalar field such as the axion, but the Einstein frame
is the one in which to calculate the spectra $\calp_\rho$ and $\calp_g$
\cite{LS3,KST}.

The original version of extended inflation was based on a Jordan--Brans--Dicke
theory, characterised by a mass scale $M$, and the Brans--Dicke parameter
$\omega$. It was quickly ruled out \cite{W,LSB}, because the upper limit on
$\omega$ during inflation, to be discussed in a minute, is far below the value
required at present by time-delay tests of general relativity \cite{REAS}. A
succession of improved models has been provided to circumvent this difficulty
\cite{LSB,HK,BM,SA}. Many of these leave inflation unchanged, merely shielding
$\omega$ from present day observation by invoking some mechanism to make the
Einstein and Jordan frames coincide quickly after inflation ends. Such models
include those where a potential is introduced for the Brans--Dicke field
\cite{LSB}, amongst others \cite{HK}. Let us consider them first.

In the Einstein frame they correspond to power-law inflation with
$n=(2\omega-9)/(2\omega-1)$ and $V_2\quarter=M$. An upper bound on $n$ follows
from the fact that the effect of bubbles forming during the early stages of
inflation has not been seen in the microwave sky. It is taken from a detailed
calculation by Liddle and Wands \cite{LW1}, following heuristic analyses by
several authors \cite{W,LSB}. In \cite{LW1} the initial bubble spectrum was
calculated, incorporating evolution from the end of inflation to the last
scattering surface and the expected anisotropy in specific experiments was
analysed. The constraint arose from a combination of experimental sky coverage
and the ability to resolve small bubbles, and the most efficient constraint
was found to come from the then-current COBE results based on the first six
months of observations \cite{S91}, assuming pixel-to-pixel variations of no
more than $10^{-4}$. In fact, it is not clear that the new COBE data have much
improved this, as the signal is a few ($\sim 3$) bright spots somewhere on the
microwave sky, and they may be susceptible to shielding by the galaxy or
misidentification as sources in the data reduction. In any case, the
constraint is only a weak function of the void resolvability, so we conclude
the constraint given then is still the best available, while noting that it is
likely to be extremely conservative, as discussed in \cite{LW1}.

The bubble constraint is plotted in Figure~2, assuming that the dark matter is
cold. It depends weakly on the matter content of the universe, and for
comparison the effect of taking instead pure hot dark matter is also shown. As
at most a small admixture of hot dark matter is allowed cosmologically, the
true result is nearer to the cold dark matter result. Also, with the addition
of any hot dark matter free-streaming removes yet further short-scale power
relative to the large-scale power detected by COBE to which we are
normalising, which makes the velocity data even harder to reconcile. Combined
with the constraint on $M$ one sees that $n \lsim .75$, which rules out these
models.

More general models, in which $\omega$ becomes dynamical, have to be examined
in their own light. For example, where one simply sets $\omega = \omega(\Phi)$
\cite{BM}, it has been shown \cite{LW2} that the bubble constraint is
substantially stronger than one's naive expectation that the normal constraint
on $\omega$ would apply at the epoch where the large bubbles are formed. On
the other hand, the hyperextended model \cite{SA} can be arranged such that
the phase transition completes via the expansion becoming subluminal rather
than by an increasing nucleation rate. It thus evades big bubbles in an
unusual way, but detailed calculations extending \cite{LW2} would have to be
carried out to investigate whether the spectrum of density fluctuations could
be made flat enough, and the gravitational wave contribution sufficiently
small, that a satisfactory model could be arranged.

{\em Note added in proof:} Since completing this work, we have received three
further preprints \cite{THREE} discussing the possible role of gravitational
waves on the microwave background. Where we overlap, our results are in good
agreement. None of these papers however exhibits explicit constraints on
power-law models, or discusses the viability of extended inflation in that
light. Finally, we have also received a preprint from Crittenden and
Steinhardt \cite{CS} which addresses many of the questions concerning
hyperextended inflation which we raised in our final paragraph.

\section*{Acknowledgements}

We thank Ed Copeland, Rocky Kolb, Josh Frieman, Jim Lidsey, David Salopek,
Paul Steinhardt, Ewan Stewart and Andy Taylor for helpful discussions and
comments. ARL is supported by the SERC, and acknowledges the use of the
STARLINK computer system at the University of Sussex.

\newpage
\section*{Figure Captions}

{\em Figure 1. Constraining the bias.}\\
Spectral slope limits for power-law and extended inflation, based on results
from COBE and from IRAS, at $2$-sigma exclusion both above and below. The COBE
bound includes the effect of gravitational waves. We take the limit as $n >
0.84$; however, note the sensitivity of the limit if the true fluctuations are
well below the COBE upper limit.

\vspace{0.2cm}
\vspace*{1em}\noindent
{\em Figure 2. Extended inflation parameter space.}\\
This figure shows the allowed regions of the $\omega$-$M$ parameter plane for
extended inflation, in the light of limits on big bubbles and on microwave
fluctuations. The inflaton potential at the end of inflation is $M^4$, and the
spectral index is $n=(2\omega-9)/(2\omega-1)$. The microwave constraint
applies also to power-law inflation.
\end{document}